# Understanding and exploiting interfacial interactions between phosphonic acid functional groups and co-evaporated perovskites

Thomas Feeney,[1,2,9] Julian Petry,[1,2,9] Abderrezak Torche,[3,4] Dirk Hauschild,[5,6,7] Benjamin Hacene,[1] Constantin Wansorra,[5] Alexander Diercks,[1] Michelle Ernst,[3,4,8] Lothar Weinhardt,[5,6,7] Clemens Heske,[5,6,7] Ganna Gryn'ova,[3,4] Ulrich W. Paetzold,[1,2] and Paul Fassl[1,2,10]

1. Light Technology Institute, Karlsruhe Institute of Technology (KIT), 76131 Karlsruhe, Germany

2. Institute of Microstructure Technology, Karlsruhe Institute of Technology (KIT), 76344 Eggenstein-Leopoldshafen, Germany

3. Heidelberg Institute for Theoretical Studies (HITS gGmbH), Heidelberg University, 69115 Heidelberg, Germany

4. Interdisciplinary Center for Scientific Computing (IWR), Heidelberg University, 69115 Heidelberg, Germany

5. Institute for Photon Science and Synchrotron Radiation (IPS), Karlsruhe Institute of Technology (KIT), 76344 Eggenstein-Leopoldshafen, Germany

6. Institute for Chemical Technology and Polymer Chemistry (ITCP), Karlsruhe Institute of Technology (KIT), 76131 Karlsruhe, Germany

7. Department of Chemistry and Biochemistry, University of Nevada, Las Vegas (UNLV), Las Vegas, Nevada 89154-4003, USA

8. Department of Chemistry, University of Zurich, 8057 Zürich, Switzerland

9. These authors contributed equally

10. Lead contact

*Correspondence: paul.fassl@kit.edu

*Correspondence: ulrich.paetzold@kit.edu

*Correspondence: ganna.grynova@h-its.org

## Summary

Interfacial engineering has fueled recent development of p-i-n perovskite solar cells (PSCs), with self-assembled monolayer-based hole transport layers (SAM-HTLs) enabling almost lossless contacts for solution-processed PSCs, resulting in the highest achieved power conversion efficiency (PCE) to date. Substrate interfaces are particularly crucial for the growth and quality of co-evaporated PSCs. However, adoption of SAM-HTLs for co-evaporated perovskite absorbers is complicated by the underexplored interaction of such perovskites with phosphonic acid functional groups. In this work, we highlight how exposed phosphonic acid functional groups impact the initial phase and final bulk crystal structures of co-evaporated perovskites and their resultant PCE. The explored surface interaction is mediated by hydrogen bonding with interfacial iodine, leading to increased formamidinium iodide (FAI) adsorption, persistent changes in perovskite structure, and stabilization of bulk α-FAPbI$_3$, hypothesized as due to kinetic trapping. Our results highlight the potential of exploiting substrates to increase control of co-evaporated perovskite growth.

## Introduction

Within the realm of emerging semiconductors for next-generation photovoltaic (PV) devices, organic-inorganic metal halide perovskite solar cells (PSCs) are promising.[1] They owe this to their tunable bandgap, coupled with high defect tolerance, absorption coefficients,[2] and carrier diffusion lengths.[3] While current certified power conversion efficiencies (PCEs) of planar inverted p-i-n PSCs lag behind their n-i-p counterparts,[4,5] the former possess several inherent advantages. These include their straightforward incorporation into monolithic tandem PV devices,[6,7] low temperature fabrication requirements, minimal current-voltage hysteresis, and high operational stability.[4,8–11]

Absorber fabrication methods for PSCs can be roughly divided into two classes, i.e., solution processing and vacuum-based thermal evaporation.[12] Solution processing is the most ubiquitous method, encompassing spin coating, inkjet printing, slot-die coating, and dip coating.[13] Such techniques allow simple process optimization, require inexpensive equipment, and enable rapid absorber deposition. An alternative class of deposition methods to fabricate high quality perovskite thin films is vacuum-based thermal evaporation (hereafter referred to as evaporation). Compared to solution processing, evaporation is well suited to deposit on textured surfaces and readily maintains homogeneity over larger areas – leading to reduced upscaling losses for industrially relevant active areas.[14–16] Evaporated perovskites have seen a recent expansion into compositions traditionally only formed through solution processing, including triple-cation,[17] methylammonium (MA)-free, and wide bandgap compositions.[18,19] Among these options, perovskites utilizing formamidinium (FA) halides as sole organic cations are particularly promising due to the expected improved performance and stability.[18]

The growth of co-evaporated perovskites shows a far greater dependence on a suitable choice of substrate compared to other fabrication methods, requiring optimized evaporation rates for good crystallization and device performance.[20–27] Olthof *et al*. were the first to report that perovskite formation during co-evaporation exhibits a strong surface-dependence that affects both film morphology and band alignment.[21] Many studies sought to understand this behavior, for example postulating catalytic decomposition for certain surfaces (such as TiO$_2$).[21,22] However, this is not sufficient to explain observed differences between catalytically inactive substrate materials. Abzieher *et al.* attempted a more universal explanation, observing changes in initial perovskite growth through suppression or incitement of Oswald ripening caused by substrate surface properties.[23] For inferior surfaces that result in non-optimal film morphology and low device performance, the initial formation of numerous small Pb-rich grains was observed, which tend to have reduced rates of organic cation incorporation.[22,23] A correlation was drawn between the substrate surface polarity (as determined by the water contact angle) and preferential initial crystal growth, highlighting that substrate-dependent initial growth conditions strongly influence bulk material properties. These observations were recently corroborated by Yan *et al.*, who showed that co-evaporated perovskite film properties can be controlled on a variety of substrates by an ultrathin perovskite template layer deposited *via* prior separate sequential evaporation.[20]

Critically, surface-based functional group interactions add a further layer of complexity for surface-dependent growth, the clearest example being the co-evaporation of $FAPbI_3$-based absorbers on MeO-2PACz reported by Ross *et al.*[24] MeO-2PACz is a member of the nPACz family – a group of self-assembled monolayer based hole-transporting layers (SAM-HTLs) that offer near-lossless contacts. They are comprised of a carbazole functional body with a phosphonic acid anchoring group.[7,28] Substantial differences in crystal growth of co-evaporated perovskites were observed, depending on whether or not MeO-2PACz was washed with ethanol to remove residual bulk SAM material before the deposition.[24] As the difference in contact angle for washed and unwashed MeO-2PACz is insignificant, as shown in our previous work,[29] the authors postulated that the difference in growth can be attributed to the formation of hydrogen bonds between exposed phosphonic acid functional groups and FAI during the deposition process. Hydrogen bonding between $FA^+$ cations and the substrate material has been extensively reported.[24,30,31] It has previously been attributed to a reduced formation energy for alpha (α)-phase $FAPbI_3$,[24,30,31] as well as an increased defect formation energy and enhanced thermal stability, even for solution-processed perovskites.[30,32] Understanding of these interfacial mechanisms and their relation to evolving structural properties during perovskite crystal growth *via* co-evaporation is currently limited.

Recently, our group introduced deposition of nPACz SAM-HTLs of controlled thickness *via* thermal evaporation from a crucible in vacuum.[29] In the present work, we employ a combination of this approach and standard solution-processed deposition of SAM-HTLs to perform an in-depth analysis of the impact of exposed phosphonic acid functional groups on co-evaporated perovskite film formation and whether it is mediated through interfacial hydrogen bonding. The presence of exposed phosphonic acid functional groups is controlled by washing the SAM layers with ethanol to remove unbound molecules. We demonstrate the immediate effect of the surface-dependent growth by comparing the maximum possible PCE of our co-evaporated PSCs on hydrogen bonding (non-washed) substrates with their non-bonding (washed) equivalents. Bulk properties and interfacial losses are explored *via* a combination of PV parameter analysis, x-ray diffraction (XRD), and photoluminescence quantum yield (PLQY) measurements. Critically, the presence of phosphonic acid functional groups has a substantial impact on the incorporation rate of organic cations, which strongly shifts the required FAI evaporation rate for optimal device performance. The presence of phosphonic acid functional groups appears beneficial for crystal growth, as it results in a preference for columnar crystal growth and effectively suppresses the formation of the photoinactive delta (δ)-FAPbI$_3$ phase during co-evaporation, which enables reproducible room temperature formation of α-FAPbI$_3$.

We explore the hydrogen bonding between various organic cation materials and common SAM-HTLs *via* solution-based nuclear magnetic resonance (NMR). Furthermore, density functional theory (DFT) analysis of various SAM/perovskite combinations allows us to identify the chemical and physical causes for our observations. We find hydrogen bonding between surface iodine of the FAPbI$_3$ layer and phosphonic acid functional groups of nPACz that are of equal strength for α-FAPbI$_3$ and δ-FAPbI$_3$, indicating that such bonds are not directly responsible for the observed change in preferential growth to the α-phase. Instead, we postulate a form of kinetic trapping, where hydrogen bonding provides an energetic barrier for the conversion of α-FAPbI$_3$ to δ-FAPbI$_3$. We experimentally support this hypothesis through x-ray emission spectroscopy (XES) and XRD measurements of thin co-evaporated perovskite and FAI films grown on 2PACz SAM-HTLs with and without exposed phosphonic acid functional groups.

This work elucidates the substrate impact on co-evaporated perovskite crystal formation and provides an explanation for how substrate-based phosphonic acid interactions lead to changes in organic incorporation rate and crystalline phase formation due to the presence of hydrogen bonding groups. In view of the need to develop reproducible and scalable fabrication methods, these results have important implications for future choices of substrate materials in evaporated perovskite-based single-junction and tandem devices.

## Results and discussion

*Performance comparison for evaporated and solution-processed nPACz layers*

Typically, 2PACz and MeO-2PACz are integrated into p-i-n PSCs by solution-based methods.[7,33] Only recently, Farag, Feeney *et al*. reported on the deposition of nPACz SAM-HTLs *via* thermal evaporation from a crucible in vacuum.[29] However, the above-discussed substrate influence on perovskite film formation makes their incorporation into all-evaporated PSC device stacks non-trivial. To evaluate the impact of SAM-HTLs with exposed phosphonic acids in co-evaporated p-i-n PSCs, we start by comparing maximum achievable device performances employing various 2PACz layers. Since differences have been previously observed for ethanol-washed and unwashed spin-coated MeO-2PACz layers,[24] we directly compare (i) evaporated unwashed, (ii) evaporated washed, (iii) spin-coated unwashed, and (iv) spin-coated washed SAM-HTLs, as seen in Figure S1. In this comparison, we employ a double-cation perovskite absorber with the composition Cs$_{0.13}$FA$_{0.87}$Pb(I$_{0.95}$Cl$_{0.05}$)$_3$ in the inverted p-i-n architecture: glass/indium tin oxide (ITO)/SAM-HTL/perovskite/C$_{60}$/SnO$_x$/Au. This composition is based on work by Lohmann *et al*., who performed bulk passivation of co-evaporated perovskites *via* incorporation of PbCl$_2$ employing four deposition sources.[34] To achieve the best PCE, the relative evaporation rate of FAI was optimized for each of the above SAM-HTL variations. More details can be found in the Experimental Procedures.

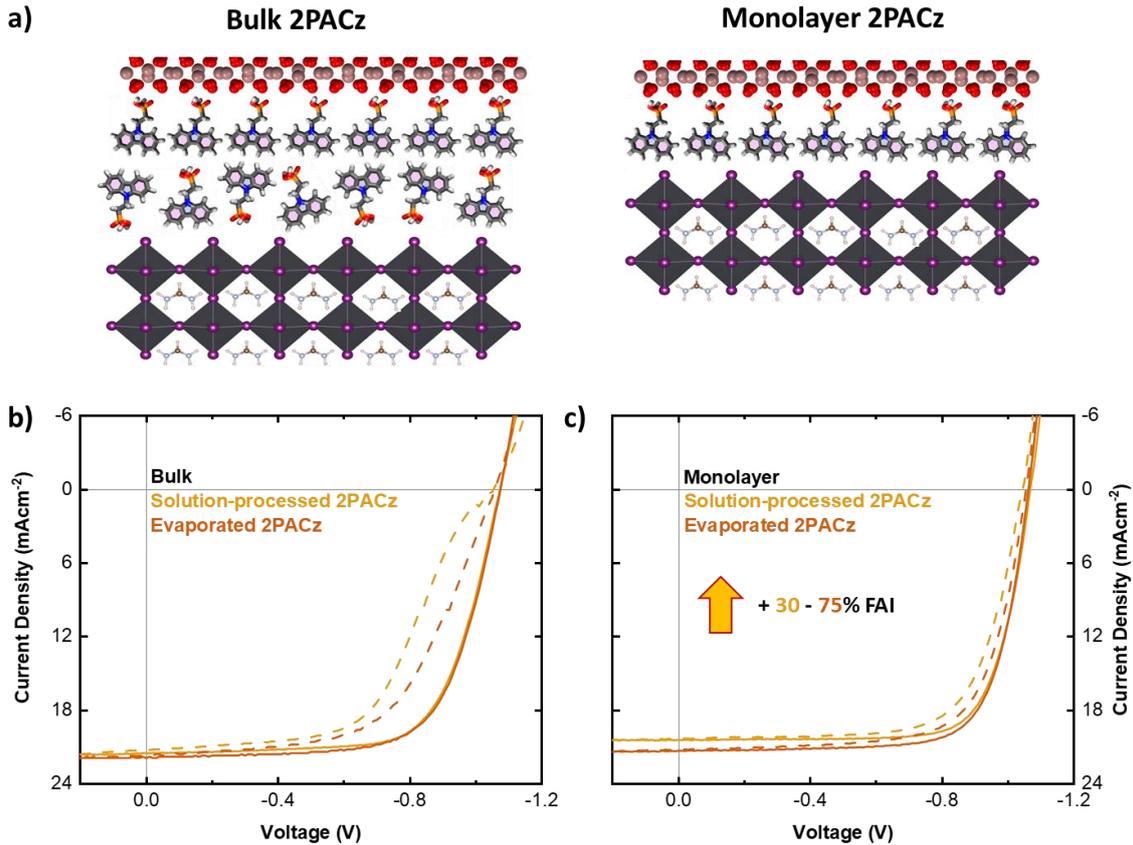

**Figure 1: Visualizations of potential SAM/perovskite interactions and resultant *J-V* curves of best-performing PSCs.** a) Schematic of perovskite grown on bulk (left) and monolayer (right) 2PACz. b) and c) *J-V* curves of the best-performing PSCs in reverse (solid) and forward (dashed) scan direction with evaporated (orange) and solution-processed (yellow) 2PACz: a) bulk SAMs and c) monolayer SAMs, including a rough estimate of the required difference in FAI rate (+30% for solution-processed and 75% for evaporated 2PACz, respectively) to achieve maximum PCE.

All solution-processed SAM-HTLs utilised high concentrations to guarantee a bulk component,[24] while evaporated SAM-HTLs featured a thickness of 4 nm, measured by a quartz-crystal microbalance (QCM). As unbound 2PACz has an approximate molecular length of 1.3 nm, it is expected that evaporated nPACz SAM-HTL layers with a thickness greater than a single molecule will exhibit a bulk component composed of unbound nPACz with exposed phosphonic acid functional groups. We expect a similar (but less controlled) bulk component in the case of solution-processed unwashed SAM-HTL layers. Bulk nPACz molecules are bound to the monolayer by Van der Waals forces. π-π interactions between carbazole functional groups are particularly strong non-covalent bonds, which are expected to lead to the formation of a self-assembled bilayer with exposed phosphonic acid functional groups.[35] The strength of these interactions is dependent on substituents, and, hence, the density of exposed phosphonic functional groups may differ between various nPACz molecules.[36,37] We note that this disordered bulk SAM layer is removed in part or even fully during standard solution-processed perovskite deposition, as observed in our previous work, where an extra washing step for a 6 nm evaporated 2PACz film before perovskite deposition did not meaningfully impact PV parameters or PLQY compared to an unwashed film.[29] Furthermore, increasing the evaporated 2PACz thickness above that of the highest PCE (i.e., from 6 nm to

20 nm) yielded only minor changes in PLQY when solution-processed perovskite layers were deposited on top. However, evaporated perovskites do not disturb this bulk SAM layer, preserving exposed phosphonic acid functional groups. Therefore, the unwashed SAM-HTL layers are expected to impact device performance and perovskite film formation. A schematic is shown in Figure 1a for perovskite films grown on 'bulk' and 'monolayer' 2PACz, respectively. Hereafter, we refer to unwashed SAM layers with the presence of exposed phosphonic acids as bulk films, and to washed SAM layers without exposed phosphonic acids as monolayer films.

Figure 1b and 1c provide current density-voltage (*J-V*) curves of the best-performing devices with each 2PACz variant. PSCs with monolayer solution-processed 2PACz achieve a maximum PCE of 17.0% with a fill factor (*FF*) of 0.77, an open-circuit voltage ($V_{OC}$) of 1.069 V, and a short-circuit current density ($J_{SC}$) of 20.6 mA cm$^{-2}$ in the reverse scan direction, while PSCs with monolayer evaporated 2PACz achieve a comparable PCE: 16.7%, *FF*: 0.74, $V_{OC}$: 1.064 V, and $J_{SC}$: 21.4 mA cm$^{-2}$. PSCs with bulk films gave PV parameters of PCE: 16.2%, *FF*: 0.70, $V_{OC}$: 1.074 V, and $J_{SC}$: 21.5 mA cm$^{-2}$ for solution-processed 2PACz, and PCE: 16.0%, *FF*: 0.69, $V_{OC}$: 1.065 V, and $J_{SC}$: 21.8 mA cm$^{-2}$ for evaporated 2PACz, respectively. Full statistics, EQE, reflectance, dark *J-V*, and maximum power point (MPP) measurements are provided in the Supplemental Information (SI), along with data sets for optimized PSCs using MeO-2PACz and Me-4PACz as a SAM-HTL (Figures S3 and S4). The integrated $J_{SC}$'s derived from EQE show currents of 20.2 mA cm$^{-2}$ and 20.7 mA cm$^{-2}$ for monolayer and 21.6 mA cm$^{-2}$ and 22.1 mA cm$^{-2}$ for bulk 2PACz from evaporated and solution processed sources, respectively. These values are within 10% of the *J-V* scan values, which is considered reasonably accurate.[38] Minor differences between $J_{SC}$ from EQE and *J-V* are expected due to differences in measuring conditions. Notably, both monolayer and bulk evaporated 2PACz yield comparable maximum device performance in the reverse scan direction as compared to their solution-processed counterparts, reinforcing the suitability of evaporated 2PACz as a HTL for all-evaporated PSCs. PSCs with bulk films exhibit greater overall differences in PV parameters between evaporated and solution-processed nPACz, which we attribute to variations in the SAM layer thickness considering the lack of accurate methods to determine the thickness of ultra-thin solution-processed layers, or potential differences in the coverage and orientation of the bulk component. We assume comparable monolayer nPACz surface coverage on ITO between deposition methods due to the similar device performance and interfacial properties as also observed in our previous work.[29] However, we have no method of determining if the bulk component coverage is also as comparable between evaporated and solution-processed layers. We note that the PCE at MPP of the best PSCs with bulk films is ~1.5 - 2% absolute lower as compared to monolayer films (see Figure S3), which could be attributed to a higher series resistance and possibly enhanced magnitude of ion migration, in line with the observed significantly larger hysteresis in these samples.[39,40]

To understand potential interfacial differences for optimized co-evaporated perovskites on SAM-HTLs with and without exposed phosphonic acids, we examine the quality of the HTL/perovskite interface for the different 2PACz layers through an analysis series on the half-stack ITO/2PACz/perovskite. We refrain from introducing an electron transport layer (ETL) to exclude the known substantial non-radiative recombination losses at the perovskite/ETL interface when employing C$_{60}$.[41] For ideal bulk evaporated and solution-processed 2PACz, we obtain average implied $V_{OC}$ values of 1.088 V and 1.045 V, respectively, compared to monolayer counterparts of 1.042 V and 1.052 V (see Figure S5; respective PLQY data is shown in Figure S6). This indicates a slight change in surface recombination properties that persists independent of the deposition method. We note that the PLQY measurements are performed in ambient atmosphere, potentially explaining the slightly lower implied $V_{OC}$ values as compared to the champion PSCs shown in Figure 1. The overall comparison of bulk against monolayer MeO-2PACz and Me-4PACz shows trends of increased implied $V_{OC}$ for bulk SAM-HTLs (Figure S5). Ideality factors for optimized monolayer evaporated and solution-processed 2PACz are 1.46 and 1.45, respectively, while the bulk ideality factors are both 1.47 (see Figure S7). This indicates no significant change in the recombination mechanism at the HTL/perovskite interface when comparing bulk and monolayer SAM-HTLs.

We employed time-correlated single photon counting (TCSPC) measurements to determine charge carrier lifetimes. The decays typically fitted with either a bi-, a tri- or even a stretched exponential function, with a relatively high weighting of non-radiative Shockley-Read-Hall (SRH) lifetimes,[42] which makes decay lifetimes difficult to compare.[43] The typically observed fast initial decay is associated with monomolecular non-radiative trapping-detrapping processes, interfacial recombination and charge transfer processes into the contact layers.[43,44] While analysis of solution-processed perovskites can exclude charge transport layers to minimize these effects, substrate dependent growth of co-evaporated perovskites prevent such experiments.[23] Therefore, our discussion will focus on lifetimes obtained from a monoexponential fit of a truncated region of decay data at later times measured at low excitation fluence (initial charge carrier density of ~3·10$^{14}$ cm$^{-1}$) (see Figure S8 and Table S1).[32,41,43] These lifetimes we correlate to SRH recombination both in the bulk of the perovskite and at the SAM-HTL/perovskite interface, allowing to compare the optoelectronic quality of the various half-stacks. Lifetimes obtained from a bi-exponential fit are also provided in Table S1 as comparison. We observe similar SRH lifetimes for monolayer evaporated and solution-processed 2PACz, with averages of 277.6 ± 1.4 ns and 268.7 ± 1.5 ns, respectively. Notably, SRH lifetimes are much higher for bulk solution-processed SAMs, with 1080.9 ± 10.3 ns in case of 2PACz. We observe similar trends for MeO-2PACz and Me-4PACz. These results are in line with the previously discussed slightly higher implied $V_{OC}$ values for bulk SAMs, indicating that a thicker SAM layer more efficiently suppresses non-radiative recombination at the HTL/perovskite interface.

While PSCs with monolayer SAMs unilaterally outperform that with bulk SAMs, the process of obtaining a nearly stoichiometric composition to achieve maximal PCE reveals a critical difference between each studied HTL. With constant inorganic evaporation rates, bulk SAM-HTL layers require substantially lower FAI evaporation rates to obtain maximum PCE, with approximately 30% higher rates required for washed solution-processed layers and approximately 75% higher rates required for washed evaporated layers compared to each bulk counterpart (see Figure 1c). We attribute a potential cause for this difference between deposition methods to different densities of phosphonic acids at the surface. This could be due to the layer thickness, or potentially indicates that evaporation allows more ready π-π interactions between carbazole functional groups, and hence a greater alignment in the bulk component. Optimal PSCs shown in Figure 1 correspond to *equivalent FAI rate*s of: 1.05 Å s$^{-1}$ for monolayer solution-processed 2PACz, 1.05 Å s$^{-1}$ for monolayer evaporated 2PACz, 0.8 Å s$^{-1}$ for bulk solution-processed 2PACz, and 0.6 Å s$^{-1}$ for bulk evaporated 2PACz (see Experimental Section for more details).

As comparison, we provide *J-V* curves of respective PSCs with monolayer 2PACz utilizing equivalent FAI rates optimized for bulk 2PACz in Figure S2, clearly demonstrating strong differences and decreased performance. Literature has previously established that, especially for co-evaporated perovskites, the initial growth conditions can strongly impact the bulk. This has been shown both for the crystal structure of the bulk material, and for the organic incorporation rates.[20,23] We posit a similar mechanism is occurring in this case. We note that cursory XRD and PLQY peak position analysis did not identify a shift in (001) peak position or PLQY wavelength with respect to bulk or monolayer component, which is interpreted to mean that composition is approximately comparable between samples. We cannot exclude the possibility that the observed change in required FAI rate would result in a different stoichiometry for compositions with different halide content (i.e., Br and/or Cl), but testing this was beyond the scope of the current work. These results potentially allow higher overall deposition rates for perovskite growth without impacting composition or resulting in thermal decomposition.[45]

*Growth trends of perovskites on various nPACz substrate configurations*

Expanding the trend observed for 2PACz to other nPACz materials (MeO-2PACz and Me-4PACz) reveals that all solution-processed bulk nPACz layers require approximately 30-40% lower FAI rates to achieve maximum PCE compared to their monolayer counterparts, an extreme example of substrate-dependent growth. However, previous reports on substrate-dependent growth did not find substantial changes in the required process window,[23,24] while our findings show the need for large changes in rate even for chemically very similar substrate materials. To clearly demonstrate the large shifts in optimal FAI deposition rate, a stoichiometry series was undertaken for monolayer and bulk 2PACz deposited *via* spin coating. For a set perovskite composition, $Cs_{0.13}FA_{0.87}Pb(I_{0.95}Cl_{0.05})_3$, the *equivalent FAI rate* was varied to identify the optimal performance with respect to PV parameters, as shown in Figure 2. Samples were characterized with *J-V* measurements to determine trends in device performance, and XRD was performed on unannealed and annealed perovskite films that gave the highest PV performance to explore differences in crystal growth on various SAM-HTLs. Full statistics and a truncated series employing evaporated SAM-HTLs can be found in the SI, Figures S9 - S12.

One clear observation is a rather sudden drop in PCE at a certain equivalent FAI rate, primarily induced by a substantive drop in $J_{SC}$, for most SAM-HTLs (see Figure 2a, Figures S9 - S12). This is associated with excess organic cations in the perovskite film,[24,46,47] and the presence of this drop is therefore attributed to a change in the required FAI rate for ideal stoichiometry. This drop is specifically reached for all bulk SAMs, which exhibit a rather small processing window for high PCE, while, for monolayer SAMs, the processing window is larger and only the highest studied FAI rate shows a strong drop in PCE (e.g., for 2PACz and MeO-2PACz).

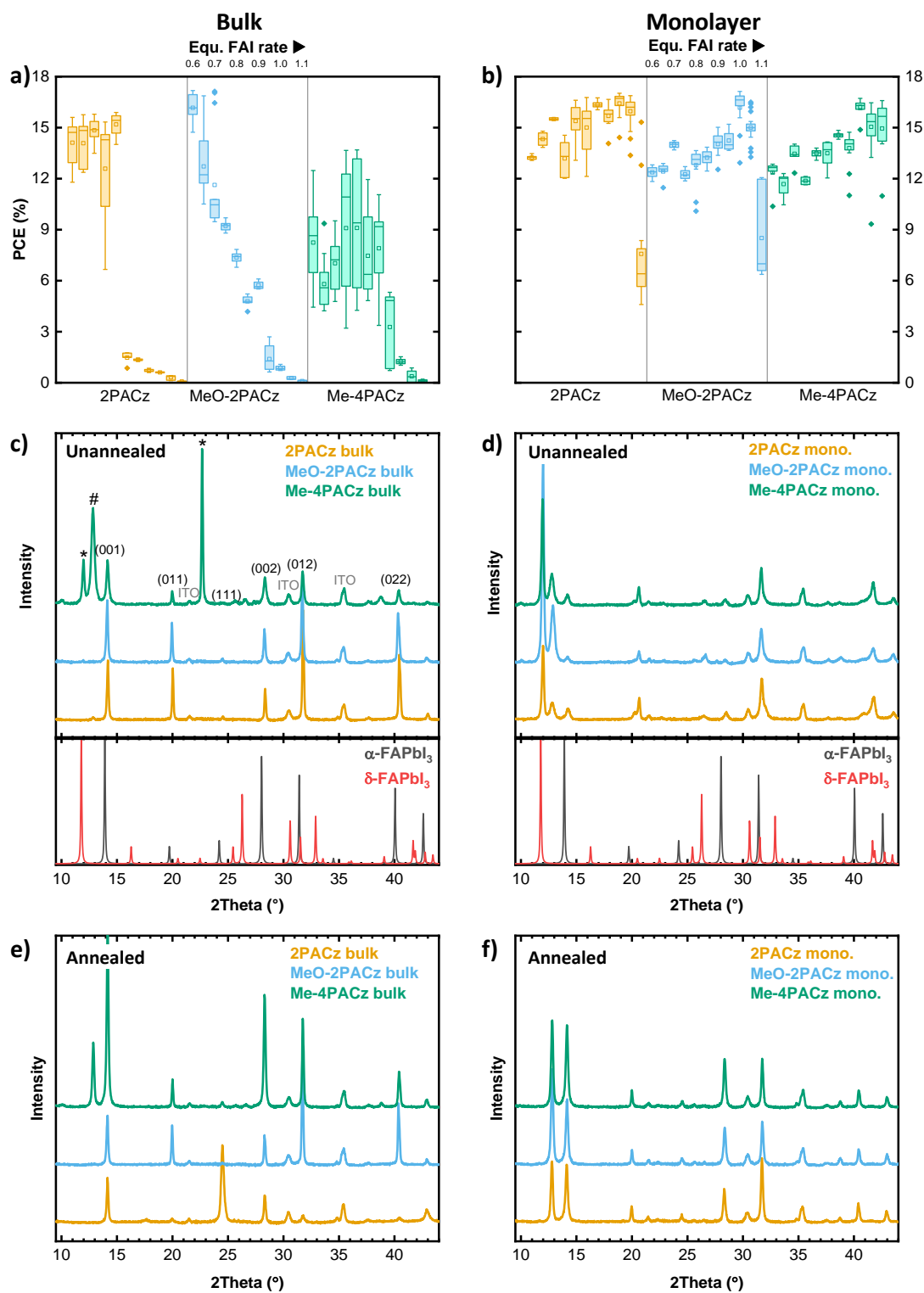

**Figure 2: Device performance and crystal structure changes between materials and deposition methods.** A) and b) Statistical trends of PCE for p-i-n PSCs with varying rates of FAI for solution-processed nPACz as SAM-HTLs, a) as deposited to leave residual bulk material and b) with a washing step to ensure the presence of a monolayer. For each series, the equivalent FAI rate is varied from 0.6 Å s$^{-1}$ to 1.1 Å s$^{-1}$ in 0.05 Å s$^{-1}$ steps. Only the reverse scan is shown. sample number ranged from 4 to 16, with an average of 10 per condition. The only exception to this range was 0.7 Å s$^{-1}$ monolayer 2PACz, which has 3 samples. Relevant FF, $J_{SC}$ and $V_{OC}$ are present in Figures S9 and S10. XRD of unannealed (c, d) and annealed (e, f) films, deposited at the same stoichiometry

as the best-performing PSCs with bulk (c, e) and monolayer (d, f) solution-processed nPACz. In c), δ-FAPbI$_3$ peak locations are labelled with *, PbI$_2$ peak locations with #, ITO peaks, and important perovskite orientations are also given. We assign the Me-4PACz peak at 22.6° to δ-FAPbI$_3$, the over pronounced height of the peak is assumed due to the highly irregular surface coverage (see Figure S13).

While our perovskite film takes the form Cs$_{0.13}$FA$_{0.87}$Pb(I$_{0.95}$Cl$_{0.05}$)$_3$, in our discussion of observed crystal phases, for simplicity, we refer to it as α-FAPbI$_3$ and δ-FAPbI$_3$ as the inclusion of Cs and Cl do not prevent the formation of phases analogous to pure FA-perovskites. For sufficiently low Cs or Cl content, as in our perovskite, the inclusion of such materials will, for example, lead to a minor shift in peak position (i.e., the (001) peak shifts from 13.98° to 14.08° for 10% Cs), but not to peak splitting or other meaningful changes.[48–50] Comparing peak positions of (001) peaks for PbI$_2$, α-FAPbI$_3$ and δ-FAPbI$_3$ does not reveal significant changes between samples. We observe a shift from an expected position from the ISCD FAPbI$_3$ reference of 13.9° to an average of 14.16° for the α-phase and from 13.78° to 13.98° for the δ-phase of our perovskite. We attribute these results to differences in measurement setup or the different stoichiometry of our samples compared to the FAPbI$_3$ reference measurement.[48,50]

Comparing monolayer and bulk nPACz, the XRD patterns of glass/ITO/perovskite half-stacks displays three noteworthy changes. First, a strong prevalence of δ-FAPbI$_3$ XRD peaks (@ 11.79°) was observed for unannealed perovskite films grown on monolayer nPACz films (Figure 2d), which is not present in unannealed films grown on bulk 2PACz and MeO-2PACz. The latter exclusively exhibit α-FAPbI$_3$ XRD peaks, indicating room temperature formation of α-FAPbI$_3$. An exception to this trend is observed for Me-4PACz; however, this can readily be attributed to uneven film coverage, which has been previously observed for solution-processed Me-4PACz layers.[51–55] Poor wettability has been theorized to be caused by the formation of SAM micelles, which require lower concentrations of SAM materials with longer alkyl chains.[55] This is further indicated by the fact that perovskites co-evaporated on bulk solution-processed Me-4PACz display dot-like coverage that is not present in equivalent perovskites on bulk evaporated Me-4PACz (Figure S13), an effect that is reduced for washed monolayer Me-4PACz films. The second noteworthy XRD change is the significantly reduced presence of crystalline PbI$_2$ in annealed films in case of bulk SAMs. The presence of significant PbI$_2$ peaks for monolayer films potentially indicates that perovskite film growth was non-ideal in terms of organic cation adsorption. Uneven coverage of Me-4PACz is a potential explanation for the continued presence of a PbI$_2$ peak in bulk layers of that material. Finally, significant changes in relative XRD peak intensity are observed that are related to changes in preferential growth.[23] First, perovskites grown on bulk nPACz materials exhibit a decreased relative intensity of the (001) crystal plane compared to secondary planes (see Figure S14). Most notable ratios of peak areas compared to the (001) plane are 2.99 for the (111) plane of bulk 2PACz compared to 0.06 for the comparable monolayer, and 1.71 for the (012) plane of bulk MeO-2PACz compared to 0.56 for the comparable monolayer. In general, bulk nPACz displays significant variation in peak intensity and distribution between materials, presented in Figure S14.

As all materials were deposited using equimolar solutions, we consider this a potential indicator that the density and arrangement of exposed phosphonic acid functional groups can differ between bulk materials. We propose two mechanisms to explain the differences in observed growth. The first relates exclusively to Me-4PACz, which exhibits uneven coverage that would lead to unexpected perovskite phases as large regions of the film are potentially covered by a monolayer or bare ITO, both of which significantly impact perovskite growth. As a more general mechanism, π-π interactions between carbazole functional groups are substituent dependent,[36,37] causing the expected density of 'bilayer' nPACz materials with exposed phosphonic functional groups to differ between our materials. This is coupled with the expected orientation of phosphonic acid functional groups to become non-uniform at low coverages due to the potential for tilted SAM configurations,[28] which are more likely with increasing alkyl chain length. Changes in this density of exposed phosphonic acids may also explain the differences in relative FAI rate required for the highest performing devices. Relative peak areas of monolayer nPACz are roughly comparable for the studied peaks, indicating approximately comparable growth (see Figure S14). A high degree of similarity also indicates that the additional functional groups present on exposed carbazole for monolayer conditions do not significantly impact FAI optimal rate or perovskite growth. The comparable optimal FAI rates for monolayers also indicate similar coverage of the ITO is expected.

Previous work has attributed both minimization of a residual crystalline $PbI_2$ signal and increases in the peak intensity of crystal planes other than (001) as signifiers of a more suitable substrate for co-evaporation.[23] Specifically, they postulate that a change in relative peak intensities to favor (011), (111) and (012) compared to (001) as indicative of columnar growth.[23,24] To shed light on this, we performed cross-sectional and surface scanning electron microscopy (SEM) of completed devices with perovskite grown on bulk and monolayer 2PACz. Visual differences in grain orientation are not as self-evident as in our previous work,[23] however we still observe a slight tendency for a reduced number of vertical grain boundaries, potentially indicative of more columnar growth (see Figure S15). Average crystal grain sizes, determined from surface SEM measurements, do not show statistically significant differences between underlayers (113.7 ± 4.73 nm for bulk and 100 ± 3.2 nm for monolayer 2PACz). However, we observe some grains with a significantly higher SEM response for the monolayer sample, leading to brighter colouration, which have previously been associated with Pb-rich perovskite grains,[33] supporting the increased $PbI_2$ observed in the corresponding XRD.

To corroborate that the observed XRD changes stem from the discussed differences between monolayer and bulk SAMs prompted us to form a monolayer film with exposed phosphonic acids. Al-Ashouri *et al.* recently showed that incorporating 1,6-hexylenediphosphonic acid (6dPA) into Me-4PACz results in SAMs with exposed phosphonic acid functional groups.[53] As the present work primarily focusses on 2PACz, we replicated this effect with 1,4-butylenediphosphonic acid (4dPA). For this, washed 2PACz/4dPA films with varying molar fractions (0%, 10%, 20%, 30%, 60%) of 4dPA were prepared. Perovskites were deposited at 0.75 Å s$^{-1}$ equivalent FAI rate, which is expected, in the presence of exposed phosphonic acids, to form exclusively α-$FAPbI_3$ during co-evaporation (i.e., for unannealed perovskite films). Indeed, as shown in Figure S16, increasing the molar fraction of 4dPA directly correlates with a trend of decreasing intensity in the (001) δ-$FAPbI_3$ XRD peak and increasing intensity of the (001) α-$FAPbI_3$ XRD peak, with complete conversion observed for 60% 4dPA, confirming our expectations.

Inclusion of 4dPA into devices at all tested molar fractions resulted in reduced PCE compared to bulk or monolayer SAMs with equivalent FAI rates (see Figure S17). Due to the impact of excess $PbI_2$ on photovoltaic parameters,[47,56] comparisons of said parameters between samples are inherently not reliable, as an increase in 4dPA leads to a decrease in excess $PbI_2$ (see Figure 16), probably due to increased incorporation of organic cations. The maximum $V_{OC}$ remains steady up to 20% 4dPA, while average $V_{OC}$ decreases slightly, before a sharp decline at 30% and above. A potential explanation is increased FAI absorption due to the high density of exposed phosphonic acids leading to an excess. However, excess FAI causes a non-linear drop in $V_{OC}$ (compare Figure S9), as opposed to the linear decrease observed with respect to 4dPA concentration. We attribute the $V_{OC}$ drop (and related lower PCE compared to bulk or monolayer 2PACz) to the insulating behaviour of alkyl linkers, with any inclusion of 4dPA reducing the conductivity and hole extraction capabilities of 2PACz.

Abzieher *et al*. estimated surface polarity *via* contact angle measurements with an $H_2O$ droplet, which was correlated with preferential crystal growth during co-evaporation.[23] In our case, however, washed and unwashed 2PACz and MeO-2PACz substrates deposited *via* evaporation or spin coating displayed no consistent difference in contact angle for any of the studied materials (Figure S18), excluding it as an explanation for the observed differences in required relative organic cation rate. Therefore, we conclude that the presence of exposed phosphonic acid functional groups has a greater impact on interfacial interactions and determining perovskite growth than surface polarity. Combined, these results support the presence of an interaction between exposed phosphonic acid functional groups and the perovskite during film formation. These exposed groups are not present in the case of monolayer nPACz films, but highly likely in bulk materials.[35] This can explain the observed differences between unwashed evaporated and solution-processed films, as the effect appears dependent on the density of phosphonic acid functional groups, and was previously attributed to hydrogen bonding to $FA^+$ cations.[24] However, the significant impact on perovskite film formation and PV parameters prompts an in-depth investigation of this process, which will be conducted in the following sections.

*Presence and strength of hydrogen bonding between phosphonic acid functional groups and organic cations studied by liquid phase NMR analysis*

Previous literature indicates two dominant forms of hydrogen bonding between phosphonic acid functional groups and FAI.[24,32,57] (i) H…O hydrogen bonding between the phosphonic acid and amidine moiety from the $FA^+$, and (ii) OH…I hydrogen bonding between the phosphonic acid and iodine. To better understand the specific form of this interaction for nPACz materials, a series of liquid-state $^1$H-NMR measurements was conducted. Reference raw spectra of 2PACz and FAI dissolved in a deuterated dimethylsulphoxide (DMSO) solvent are shown in Figure 3a. Upon the introduction of equimolar 2PACz to the FAI solution, a number of notable changes can be observed in the FAI spectra as shown in Figure 3b. Protons attributed to the amidine moiety split into two doublets. This splitting is attributed to the hindered rotation of $FA^+$ amidine groups caused by the formation of a delocalized double bond planarizing the molecule, disrupting the equivalent chemical environment between amidine protons.[58] Interactions with the CH proton occur through vicinal 3J cis and trans coupling (coupling constant J$_{trans}$ = 15.2 Hz). Furthermore, the CH protons split into a triplet of a triplet (tt) (coupling constants J$_{cis}$ = 6.4 Hz and J$_{trans}$ = 15.1 Hz) due to interaction with the inequivalent amidine protons. Splitting of such NMR signals is described in literature, but its origin remains unclear, with some publications attributing it to the introduction of a Lewis acid.[59] Concurrent changes can be found within the 2PACz spectra by comparison with the reference peak for the phosphonic acid functional group at 8.80 ppm (compare Figure 3a). Here, a clear chemical shift is observed; upon introduction of FAI, the peak shifts to 6.68 ppm for a 0.5 molar ratio of 2PACz to FAI, while it increases to 9.67 ppm with increasing 2PACz concentration as shown in Figure 3c.

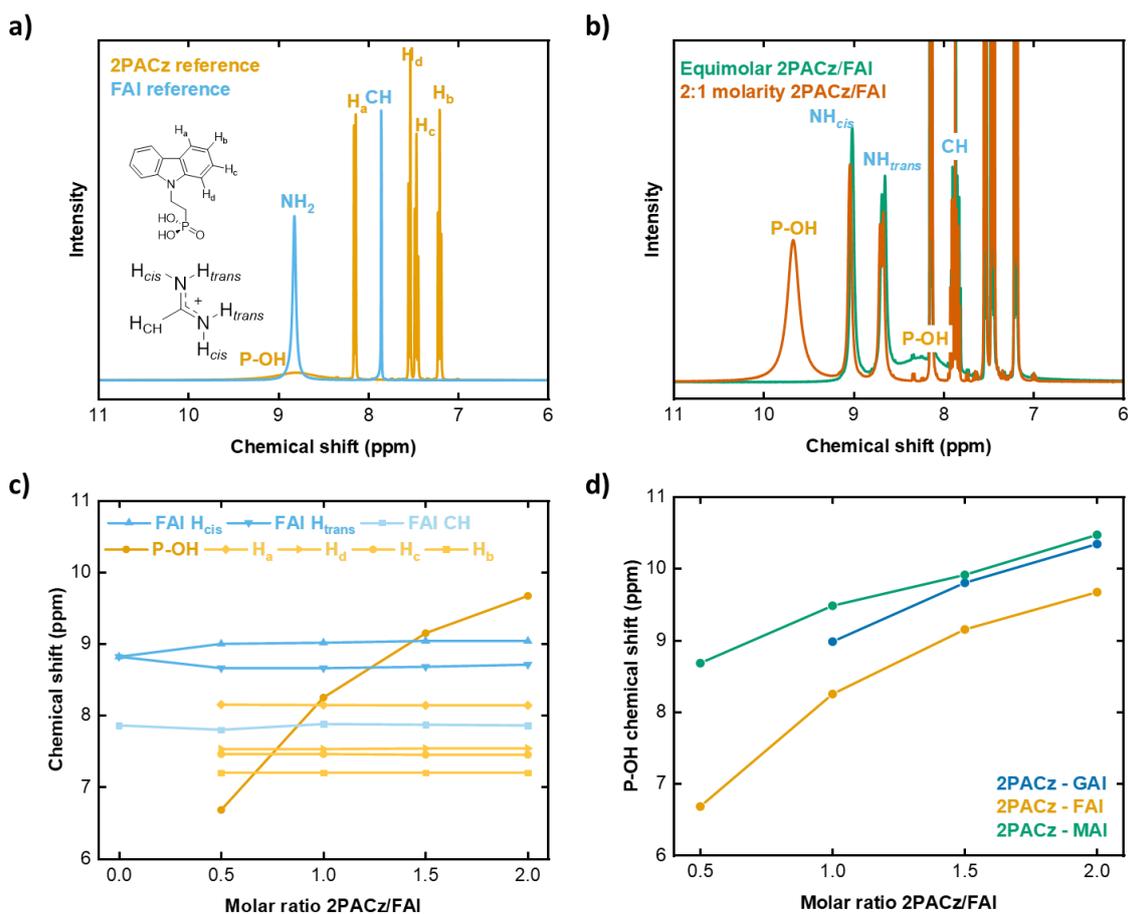

**Figure 3: NMR series identifying the form of hydrogen bonding.** a) Solution-NMR of pure FAI and pure 2PACz solutions. The inset illustrates the cis and trans variation of the deplanarized FA$^+$ molecule. b) Solution-NMR of FAI solution with two concentrations of 2PACz. c) Trend of chemical shift for each signal as a function of 2PACz concentration, indicating no shift in FAI NMR peak locations with increasing 2PACz concentration. d) Chemical shift trends of P-OH peak for FAI, MAI, and GAI with varying concentration of 2PACz. Relevant NMR spectra corresponding to the stated shifts are shown in Figure S19.

To determine which form of hydrogen bonding is present and to derive a material-dependent trend for bonding strength, several liquid phase series were conducted. These series investigate interactions between common organic perovskite cations (MAI, FAI, guanidinium iodide (GAI)), with 2PACz as a constant SAM-HTL, as shown in Figure 3d. A second series comparing interactions between FAI with two phosphonic acid containing SAM-HTLs (2PACz, MeO-2PACz) is shown in the SI (Figures S19 and S20), along with all NMR spectra. Measurements with Me-4PACz were attempted, but due to peak broadening and a relatively low peak intensity, the potential error in determining the peak position is too substantial for meaningful analysis, thus the trend of phosphonic acid peak shift of Me-4PACz series was excluded in our analysis (raw spectra are shown in Figure S21). Organic cation concentrations were maintained at 0.13 M, while 2PACz concentration was varied from 0 M to 0.26 M. A slight shift of the phosphonic acid signal of 2PACz was observed when changing the concentration in the absence of a halide source. This is attributed to interactions of phosphonic acids in non-polar and aprotic solvents such as d6-DMSO, which was employed for this study.[60] Despite a clear shift in the phosphonic acid with respect to material concentration, there is no complementary shift in the organic cation peak positions, as is evident in Figure 3c. Furthermore, peak shift magnitudes are roughly equivalent between organic cations (Figure 3d). This indicates that hydrogen bonding between the organic portion of these materials and the phosphonic acid functional groups is not observed in solution. An alternative explanation, rendered much more likely due to the equivalent shift for each cation, would be the formation of a hydrogen bond between the halide portion of these materials and the phosphonic acid.[61] This indicates that the observed interactions, and, hence, changes in optimal organic cation rate and crystallization, are potentially not exclusive to FAI. Furthermore, varying 2PACz/PbI$_2$ results in comparable peak shifts (see Figure S22), indicating that such interactions are not exclusive to organic precursors. However, the direct deposition nature and reduced substrate dependence of inorganic precursors,[23,45,62] means we do not expect such interactions to meaningfully impact growth.

It is important to note that materials capable of hydrogen bonding in inert solvents are naturally expected to shift peak positions with changing concentration, as demonstrated in Figure S23.[60] Furthermore, the peak position for reference 2PACz (8.80 ppm) being inconsistent with the broader 2PACz/FAI trend in Figure 3c and 3d indicates that this system is more complex than the pure 2PACz case, which we rationalize as indicative of non-comparable interaction mechanisms. Hence, we cannot precisely quantify the strength of potential hydrogen bonding between phosphonic acid functional groups and halides of organic cation precursors, and our liquid-phase results are considered qualitative rather than quantitative. However, this does not invalidate our observation that hydrogen bonding is not observed between phosphonic acid and the organic portion of these organic cation precursors. Our repeated study using MeO-2PACz and Me-4PACz (see Figure S20, 21) supports this hypothesis and indicates that the hydrogen-bonding strength is approximately equivalent between nPACz materials. Any further interactions of MeO-2PACz through the methoxy groups were excluded, as no significant shifts of the corresponding signal were observed. Consequently, in all cases, intermolecular interactions primarily stem from hydrogen bonding of free phosphonic acid functional groups with present halide ions.

*Understanding interactions between phosphonic acid functional groups and perovskites*

To gain atomistic-level insights into the interaction mode of phosphonic acid functional groups with evaporated perovskites, and to understand whether these interactions can form the basis for the experimentally observed crystal growth rate and orientation differences between each substrate, we modelled these systems using DFT. We utilized PBE-D3, which is considered highly accurate for numerous modelling applications,[63,64] including perovskites.[65] FAPbI$_3$ was chosen as a simplified perovskite model (as FA$^+$ is the dominant cation in the investigated perovskite), with 2PACz as the SAM-HTL. Additional functional groups or variations in alkyl chain length are not expected to significantly impact the studied interactions and were thus not included in our *in silico* models. Our DFT calculations consider two potential FAPbI$_3$ unit cells, the photo-inactive hexagonal phase (δ-FAPbI$_3$), and the photoactive cubic phase (α-FAPbI$_3$). Secondary photoactive phases, the orthorhombic β-FAPbI$_3$ and tetragonal γ-FAPbI$_3$ to be precise, were not investigated as their formation requires temperatures below 153 K and 93 K, respectively, which were deemed too extreme for standard perovskite film formation.[61]

The role of FA$^+$ cation orientation in the stability of α-FAPbI$_3$ and δ-FAPbI$_3$ perovskites is well established in literature.[61,66,67] Free rotation of the FA$^+$ cation within α-FAPbI$_3$ increases the entropy of this phase as temperature rises, making it the most stable phase at temperatures above ~430 K.[67] δ-FAPbI$_3$ is more stable at room temperature, due to its large unit cell and thus the lower energetic cost of encapsulating the large FA$^+$ cation. The position of the latter is fixed within the δ-FAPbI$_3$ *via* C…I and NH$_2$…I interactions, and hence the associated entropic gain is insignificant compared to the α-phase.[68] The position of the FA$^+$ cation within the cubic and hexagonal phases were optimized using DFT at the PBE level (see Experimental Procedures for computational details). Similar to the findings of Zheng *et al.*,[67] free FA$^+$ ions position themselves in the midplane of the α-FAPbI$_3$ unit cell, oriented along the plane to minimize C…I and NH$_2$…I contacts. Conversely, for δ-FAPbI$_3$, a complex arrangement is formed, featuring ribbons of coordinated PbI$_3$ and interstitial FA$^+$ with fixed orientations between parallel ribbons to minimize C…I and NH$_2$…I contacts. Encapsulation of FA$^+$ cations lowers unit cell strain, resulting in a free energy difference of 279 meV under bulk conditions. A visualization of these geometries and the associated interaction modes is present in Figure S24.

We next simulated a perovskite slab with thickness below 50 Å in the absence of any surface interactions to allow the structures to fully relax. Due to the variation in unit cell size between α-FAPbI$_3$ and δ-FAPbI$_3$, interpolation was required for intermediary thicknesses. Critically, in the absence of entropic contributions for equivalent slab thicknesses, α-FAPbI$_3$ is more stable than δ-FAPbI$_3$ in the region below 35 Å, with an energy difference of 30 meV at the lowest simulated thickness of ~16 Å (see Figure 4a). The reason for this unexpected stability is the fact that the surface energy is lower in α-FAPbI$_3$ (compared to δ-FAPbI$_3$) in the thin slab limit. Entropic contributions will further favor α-FAPbI$_3$ due to the free movement of interstitial FA$^+$, which is restricted in δ-FAPbI$_3$. These effects, estimated as ~-253 meV at 18 °C, make α-FAPbI$_3$ more stable at the early stage of the growth, and increase the thickness at which we approach the bulk condition of thermodynamically favored δ-FAPbI$_3$. These results can explain the existence of a small α-FAPbI$_3$ peak in the XRD of unannealed co-evaporated perovskite thick films grown on monolayer SAM-HTLs (Figure 2d).

Theoretical results so far suggest that the α-phase is favored at the early stages of the crystal growth, yet δ-FAPbI$_3$ is thermodynamically preferred in the bulk limit. Given that α-FAPbI$_3$ is the only crystalline phase observed experimentally in unannealed perovskite films deposited on bulk nPACz (except for Me-4PACz), we next tested the possibility that the presence of phosphonic acid promotes α-FAPbI$_3$ growth over δ-FAPbI$_3$. For three configurations of the 2PACz molecule (A – anionic, N – neutral, and T – trans-conformation), computed binding energies with FA$^+$- and Pb-terminated perovskite α-FAPbI$_3$ and δ-FAPbI$_3$ were found to be relatively similar for the two phases (see Figure 4b), with the strongest interactions occurring at distances of ~2 Å, commensurate with non-covalent interactions (such as hydrogen bonding) rather than covalent bonds. The interaction profile for a Pb-terminated perovskite was also considered, and is present in Figure S25. Computed non-covalent interaction (NCI) energies were essentially unchanged for each perovskite phase, and we observe comparable interaction strength between the 2PACz and the perovskite surface for both FA$^+$- and Pb-terminated slabs. All associated data points for these plots are present in Tables S2 – S4.

To better capture the diverse interaction modes between 2PACz and perovskite, we scanned interaction energy profiles by laterally shifting the 2PACz molecule at a fixed 2 Å distance above the 3×3 (001) perovskite surface (see Figure 5a). Maximum computed energies of 470 meV for α-FAPbI$_3$ and 720 meV for δ-FAPbI$_3$ suggest that interactions with the phosphonic acid alone cannot explain the experimentally observed preference for the α-phase. Geometries corresponding to the minima of the scanned interaction energies, shown in Figures 5b and 5c, reveal shortened contacts between the OH group of the phosphonic acid in 2PACz and surface iodine anions of the perovskite. This finding is in good agreement with our liquid phase NMR observations, which indicate that bonding primarily occurs between phosphonic acid functional groups and the halide component of organic cations, supporting the notion that such interactions are not exclusive to FA-based perovskites.

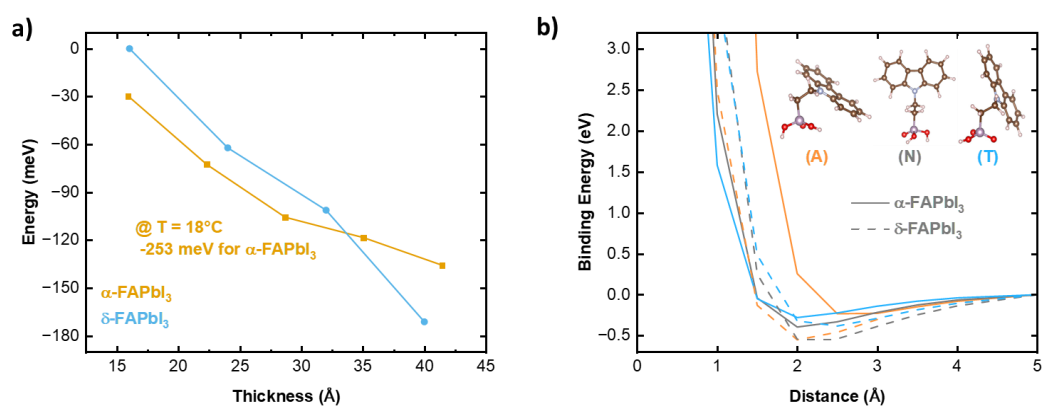

**Figure 4: DFT results of thin perovskite slabs with and without interaction energies between 2PACz and thin perovskite slabs.** a) Stability of α-FAPbI$_3$ and δ-FAPbI$_3$ slabs, determined by energy per unit cell as a function of slab thickness. These values do not include entropic contribution or surface functionalization. b) Binding energy of various 2PACz conformations as a function of the distance between substrate and material for both phases of FA$^+$-terminated FAPbI$_3$ perovskite. Inset: investigated 2PACz conformations. In these simulations, phosphonic acid is assumed to be directly exposed to the perovskite surface, corresponding to a bulk film.

These findings also provide further evidence that halide interactions are more relevant for this interface than other potential interactions such as FA$^+$ or Pb$^{2+}$.

Our results indicate that, while α-FAPbI$_3$ is thermodynamically preferred at the initial stages of film formation, δ-FAPbI$_3$ is both the dominant phase in the bulk limit and the phase more strongly stabilized by interactions with the phosphonic acid moieties of 2PACz. We therefore put forward the following mechanism explaining the experimentally observed preference for the α-FAPbI$_3$, based on the concepts of kinetic and chemical trapping (see Figure 5d). Specifically, α-FAPbI$_3$ is formed initially and preferentially to δ-FAPbI$_3$ due to kinetic (entropic) contributions outlined in Figure 4a, combined with the lowering of the surface energy outlined in Figure 5a.[66,69] Interaction with free phosphonic acid moieties introduces a passivation barrier for the conversion of α-FAPbI$_3$ into δ-FAPbI$_3$ when transitioning from thin film to bulk limit, i.e., a kind of "chemical trapping". According to Chen *et al.*,[66,69] cubic (α-FAPbI$_3$) and hexagonal (δ-FAPbI$_3$) space groups lack a simple group/subgroup connection. Transitioning from a cubic to a hexagonal phase involves complex movements and Pb–I bond breaking and forming. Therefore, such a phase transition first requires dissociation of the perovskite film from the 2PACz molecules. The height of the corresponding passivation barrier is equivalent to $N$*470 meV, where $N$ is the maximum number of phosphonic acid functional groups per 3×3 (001) surface. Interfacial passivation strategies have been previously shown to enhance the stability of α-FAPbI$_3$ and reduce the formation of δ-FAPbI$_3$ through similar surface functionalization and entropic stabilization.[68] Critically, co-evaporated perovskites grow at a relatively low rate compared to solution-processed or sequentially evaporated films, making them more vulnerable to this trapping effect.

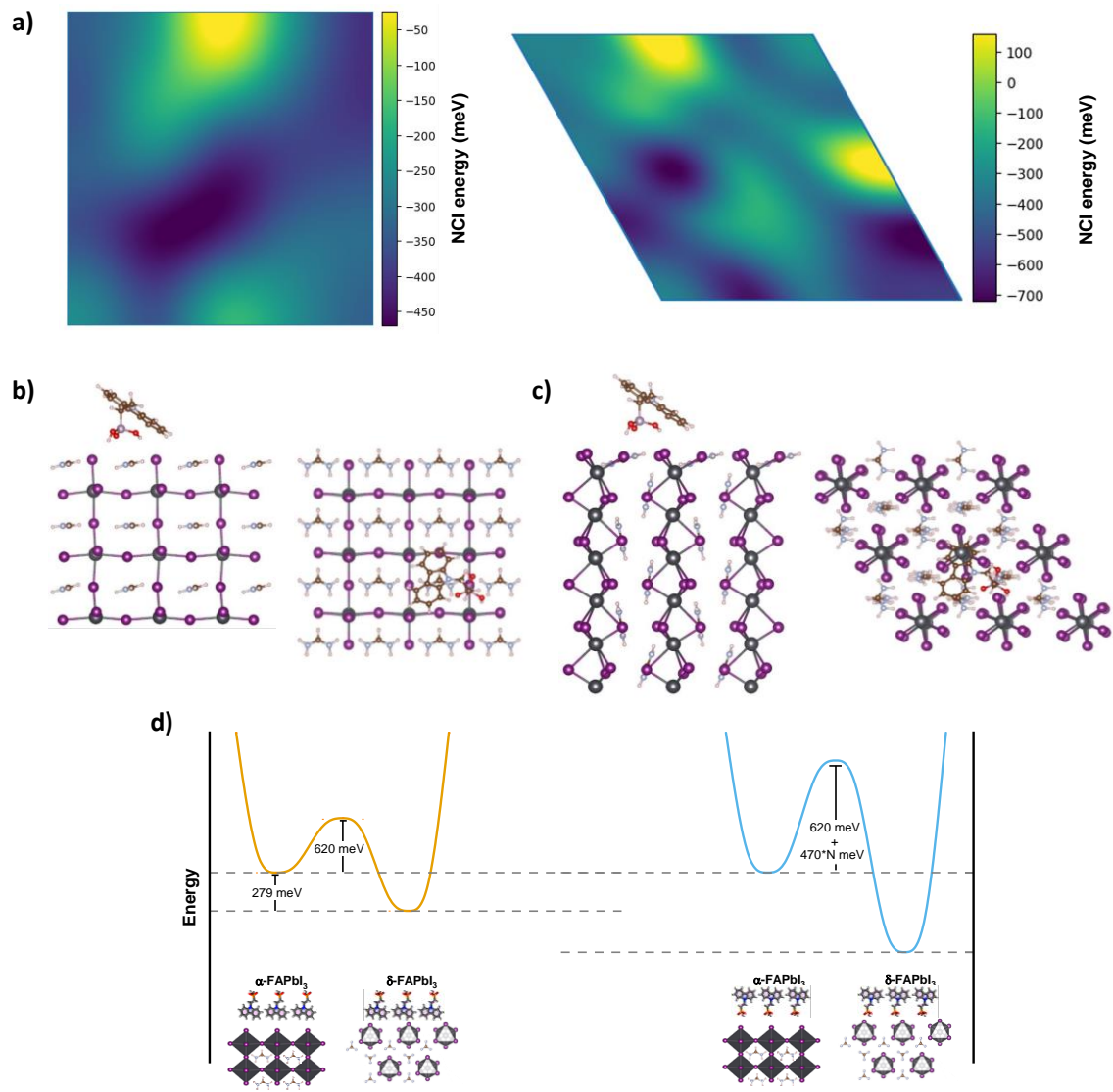

**Figure 5: DFT visualizations of interactions between nPACz and perovskites, with explanations for how such changes apply to bulk materials.** a) Computed non-covalent interaction (NCI) energies for laterally shifted 2PACz, fixed at a 2 Å distance from the surface of α-FAPbI$_3$ (left) and δ-FAPbI$_3$ (right). Geometries corresponding to the NCI energy minima of 2PACz on α-FAPbI$_3$ b) and δ-FAPbI$_3$ c) surfaces (side and top views). d) Energy diagram representation of kinetic trapping for α-FAPbI$_3$ stabilization. Activation energy of phase transitions in the absence of phosphonic acid functional groups were taken from Chen et al.[69]

To obtain a better experimental understanding of this process and investigate the impact of interactions at the interface between the SAM-HTL and the perovskite film, synchrotron-based soft x-ray emission spectroscopy (XES) was employed to determine the element-specific electronic structure at the nitrogen atoms. For this purpose, we measured non-resonant N K XES spectra (hv = 420 eV) of two different sample sets: bulk evaporated 2PACz and monolayer evaporated 2PACz. On each substrate, a ~25 nm FAI or ~25 nm perovskite film was deposited, similar to our previous work on FA-based perovskites.[70] We chose evaporation as a deposition method due to the high degree of substrate uniformity. For the bulk 2PACz substrate, we also measured the interface formation with ~25 nm PbI$_2$ and ~25 nm CsI. Additionally, the ITO substrate, a 2PACz bulk powder reference, and a bulk FAPbI$_3$ were measured, while the reference FAI spectrum was used from previous work.[71]

The ITO N K XES spectrum (Figure 6a, bottom) exhibits a broad (and relatively weak) spectral structure with a main peak at ~394 eV. While not necessarily expected in an ITO film, we suspect that this signal is due to residual nitrogen incorporated into the bulk of the ITO film, likely due to the preparation process. In contrast, the reference 2PACz spectrum displays a characteristic multi-peak structure, with the most prominent peak at ~398 eV (note that the intensity axis of this spectrum was multiplied by x0.15 to allow for better comparison of the spectral series in Fig. 6a). We attribute this peak to the transition of an electron from the highest occupied molecular orbital (HOMO), located at the nitrogen atom within the carbazole ring, into the N 1s core hole. Other peaks with lower emission energy can be associated with transitions from lower lying molecular orbitals into the N 1s core hole.

Both 2PACz substrate samples (labeled 2PACz/ITO in Fig. 6a, bottom spectra in the center panel) need to be interpreted as a superposition of the ITO N K signal (main peak at ~394 eV) and the reference 2PACz spectrum (most prominent peak at ~398 eV). While bulk 2PACz can easily be reconstructed from a combination of ITO and 2PACz reference signals, the monolayer 2PACz/ITO substrate did not require a 2PACz contribution to reproduce the signal. The former suggests that the chemical structure of the evaporated 2PACz on ITO is equivalent to the powder reference, while the latter indicates that the 2PACz signal for the monolayer film is well below that of the ITO substrate and thus corroborates the finding of a very thin layer.

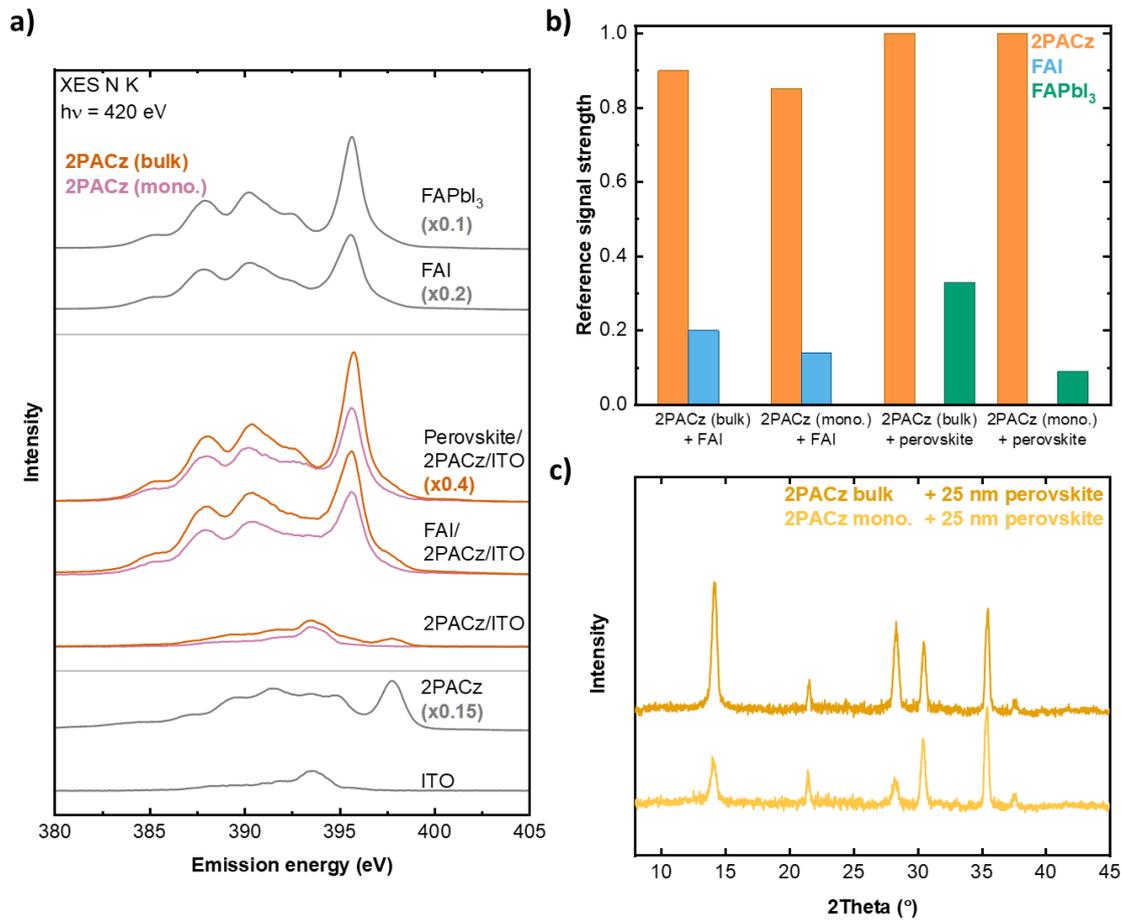

**Figure 6: XES and thin film XRD results indicating changes in perovskite and FAI deposition in thin films.** a) XES spectra of ~25 nm FAI and ~25 nm $FAPbI_3$ perovskites deposited onto bulk or monolayer 2PACz films. Reference 2PACz (pressed powder pellet), ITO, FAI,[71] and $FAPbI_3$ spectra are also shown. Where necessary, spectra were multiplied by the given truncation factor for easier viewing. b) Overview of the signal contributions of respective reference fit components to the description of the spectra in a). Note that these contributions are not precisely proportional to the nitrogen concentration due to possible differences in the photoionization cross sections of the different compounds. c) XRD spectra of unannealed ~25 nm $FA_{0.87}Cs_{0.13}PbI_3$ perovskite films deposited onto bulk or monolayer 2PACz films.

Upon FAI deposition, we observe the characteristic N K FAI spectral fingerprint (compare with the FAI reference spectra in Fig. 6a, top panel, second from top).[71] Similarly, both samples with a 25 nm $FAPbI_3$ perovskite deposition (center panel, top spectra) resemble the $FAPbI_3$ spectral fingerprint, which consists of the $FA^+$-ion and halide-derived hybrid orbitals (top panel, top spectrum, x0.1). For each pair of spectra, the bulk data show a significant higher intensity, especially for the $FAPbI_3$/2PACz/ITO samples (note the factor of x0.4). To quantify the spectral contribution, we performed a fit analysis of the spectra using the respective 2PACz substrate signal and either the FAI or $FAPbI_3$ spectrum as fitting components. For the FAI spectra, the monolayer FAI spectrum can be reproduced using [0.85 x 2PACz + 0.14 x FAI] while the bulk spectrum required a larger FAI component [0.90 x 2PACz + 0.20 x FAI]. In the same fashion, perovskite growth rates follow this trend, i.e., [1.0x 2PACz + 0.09x perovskite] and [1.0x 2PACz + 0.35x perovskite] for monolayer and bulk 2PACz, respectively. This indicates a clear increase in crystalline perovskite growth rate, partially mediated by an increased affinity of FAI to adhere to bulk 2PACz. Notably, no large spectral changes are observed for the 2PACz SAMs after deposition of 25 nm CsI or $PbI_2$ (see Figure S26). While some of the spectral features at the lower end of the valence region appear sharper after the deposition, the overall spectral shape is retained, indicating only weak (if any) interactions between these materials and the underlying 2PACz substrates.

XRD measurements of the samples with a ~25 nm thick perovskite film support the findings from the XES measurements and DFT results (see Figure 6c). Films grown on a monolayer exhibit pure α-$FAPbI_3$ growth in the thin film region (even without annealing). This is exactly what is predicted by the above-discussed DFT modelling, where α-$FAPbI_3$ is the more stable product in the thin-film region, especially when entropic effects are included. Furthermore, co-evaporated thin films grown on bulk 2PACz exhibit higher signal intensity than their monolayer counterparts, providing further evidence for the enhanced crystalline perovskite signal in the presence of exposed phosphonic acid functional groups.

Combined, the DFT, XES, and XRD results indicate a potential explanation for the lower required FAI rates to form co-evaporated perovskites on SAM-HTLs with exposed phosphonic acid functional groups. The presence of exposed phosphonic acid functional groups enhances FAI adhesion to the surface, resulting in less FAI required to form a perovskite. Increased stability of the co-evaporated perovskite, mediated by NCIs between the perovskite surface and exposed phosphonic acid functional groups, promotes the formation and stabilization of crystalline perovskite. Formation of perovskite under exposure to excess cations has also been correlated to increased formation of α-$FAPbI_3$, which will further reinforce this preferential growth.[72] Abzieher *et al*. and Yan *et al*. demonstrated that initial perovskite growth conditions readily continue to propagate during film deposition to inform bulk properties.[20,23] Hence, the initial α-phase crystalline perovskite favors rapid crystallization and exhibits increased affinity to organic cations. This initial perovskite would persist into the bulk, explaining the change in required organic cation deposition rate. With this finding, our work thus deepens the previous understanding of the complex interplay between substrate surfaces and perovskite film growth, revealing how functional groups can be a driver for initial growth conditions that radically change film formation.

**Conclusion**

This work combines novel theoretical and experimental analyses of phosphonic acid surface interactions with co-evaporated perovskite films, furthering the understanding of substrate impact on evaporated perovskite crystal formation. The presence of interfacial phosphonic acid increases the organic cation amount required to form stoichiometric films by 30-65%, depending on the employed HTL material. Furthermore, analysis of perovskite growth on various phosphonic acid containing HTLs indicates that phosphonic acids suppress δ-$FAPbI_3$ and crystalline $PbI_2$ formation, while promoting columnar growth (as indicated by a substantial change in relative XRD peak intensities). Using 4dPA as a source of exposed phosphonic acid functional groups, these observed XRD trends are replicated for monolayer films.

These studies result in photovoltaic performances of 16.2% for unwashed and 17.0% for washed solution-processed 2PACz as HTL. To the best of our knowledge, this work is the first example of co-evaporated perovskites deposited onto evaporated SAM-HTLs, which achieved 16.7% for unwashed samples and 16.0% for washed samples in case of 2PACz, reinforcing the potential for evaporation as a potential deposition method for upscaled perovskite PV. It is also the first to comparatively investigate the growth of co-evaporated perovskites on 2PACz and Me-4PACz in both evaporated and solution-processed forms.

DFT analysis of various perovskite/SAM combinations indicates that, although phosphonic acid binding interacts more strongly with $\delta$-FAPbI$_3$ than with $\alpha$-FAPbI$_3$, it serves as a trap for the $\alpha$-phase at the initial stages of crystal formation and creates a passivation barrier that inhibits conversion into $\delta$-FAPbI$_3$. Phosphonic acids strongly align to interact with interfacial iodine groups in the FAPbI$_3$ layer, an observation confirmed through liquid phase NMR of interactions between 2PACz and the common organic cations of MAI, FAI, and GAI. Our mechanistic explanation is confirmed for FA-based evaporated perovskite thin films through XES analysis of evaporated perovskite films grown on 2PACz, both with and without exposed phosphonic acid functional groups. XES indicates a greater affinity of FAI to phosphonic acid terminated surfaces, and a greater crystalline perovskite signal, which would be expected by the discovered surface passivation effect. Our findings provide a potential explanation for the shift in required FAI rates based on surface functionalization. This work therefore shows the unique capacity of co-evaporated perovskites in exploiting surface functionalization to control growth characteristics, optimal FAI rate, and device PCE, highlighting and demonstrating the potential for rational design of substrates to influence organic cation rates through surface functionalization.

## Experimental Procedures

### Resource availability

*Lead contact*

More information and requests for resources should be directed to and will be fulfilled by the lead contact, Paul Fassl (paul.fassl@kit.edu).

*Materials availability*

This study did not produce new unique materials.

*Data and code availability*

This study did not generate code. The data presented in this work are available from the lead contact upon reasonable request.

### Evaporation of self-assembled monolayer (SAM) thin films

Evaporated nPACz (2PACz, MeO-2PACz, or Me-4PACz from TCI) thin films were fabricated *via* physical vapor deposition from a crucible in a thermal evaporation system (Creaphys, OPTIvap) in a manner similar to our previous work.[29] All films were deposited at pressures of approximately $5\times10^{-6}$ mbar, with rates of 0.15-0.25 Å s$^{-1}$ measured using a quartz crystal microbalance (QCM). SAM-HTLs denoted as 'bulk films' had a final (measured) thickness of 4 nm.

**Solution-processed SAM thin films**

SAM solutions were prepared by dissolving nPACz powders (2PACz, MeO-2PACz, or Me-4PACz from TCI) in ethanol (Sigma-Aldrich, anhydrous) to create solutions with concentrations of 2.98 mM. Solutions were ultrasonicated for 30 min to equilibrate.

Solution-processed SAM-HTLs were deposited using a 1-step spin coating program (3000 rpm for 30 s) with 70 µL of solution in a nitrogen glovebox. Films were then dried at 100 °C for 10 min.

**Washing of SAM thin films**

A select subset of SAM-HTLs (labeled 'monolayer films' in this work) were washed using a 1-step spin coating program (3000 rpm for 30 s) and 150 µL of ethanol (Sigma-Aldrich, anhydrous), dynamically added after 10 s. Washed films were dried at 100 °C for a further 10 min.

**Fabrication of perovskite solar cells (PSCs)**

Planar p-i-n PSCs were fabricated with the architecture of: glass/ITO/SAM-HTL/$Cs_{0.15}FA_{0.85}Pb(I_{0.9}Cl_{0.1})_3$/$C_{60}$/$SnO_x$/Au. ITO substrates (15 Ohm/sq, Luminescence Technology) were cut to sizes of 0.16 cm × 0.16 cm and cleaned with acetone and isopropanol in an ultrasonic bath for 15 min each. Substrates were treated with oxygen plasma for 3 min immediately prior to deposition of the SAM-HTLs.

For perovskite deposition, a nitrogen glovebox integrated PEROvap (CreaPhys) evaporator was employed. Four QCMs were used to measure the rate of each material independently. A cooling inner surface, surrounding all evaporation sources, was set to -25 °C. Prior to the heating process, the system was evacuated for 60 min, with a standard base pressure at start of heating of $3 \times 10^{-6}$ mbar. For every process, the evaporation rate of each material was kept constant by manually adjusting the source temperature. Substrate temperature (18 °C) and substrate rotation speed (10 rpm) were held constant for all experiments. Inorganic rates during evaporation were: $PbI_2$ = 0.23 Å s$^{-1}$, CsI = 0.025 Å s$^{-1}$, and $PbCl_2$ = 0.01 Å s$^{-1}$. As discussed in this work, FAI incorporation and adhesion is strongly dependent on the substrate and underlying material. For this reason, the tooling factor and rates of FAI are somewhat arbitrary. The tooling factor and rates for this recipe were originally optimized for an FAI rate of 1.0 Å s$^{-1}$, which was the optimal rate (in terms of photovoltaic performance) for washed 2PACz as HTL at that time. The 'equivalent FAI rate', as shown, e.g., in Figure 2, is based on this initially optimized rate. FAI is known to exhibit decomposition at high temperatures, and is suspected to exhibit a variable tooling factor dependent on rate due to local pressure.[45] To prevent such effects over the large variation in FAI rates required, a variation in perovskite stoichiometry was achieved by keeping the FAI rate below or close to the equivalent FAI rate of ~1 Å s$^{-1}$. For a standard co-evaporation of a 550 nm perovskite layer, deposition time was approximately 180 min. All samples were annealed at 140 °C for 10 min in nitrogen atmosphere at ambient pressure.

An ETL, comprised of 20 nm fullerene ($C_{60}$, Alfa Aesar), was thermally evaporated at a 0.1 - 0.2 Å s$^{-1}$ rate under a pressure of approximately $1 \times 10^{-6}$ mbar in a vacuum evaporation system

(CreaPhys, OPTIvap). 35 nm $SnO_x$ was deposited *via* atomic layer deposition, as described in our previous work.[73] All samples were completed with the evaporation of a 75 nm Au rear electrode, with an active area of 10.5 mm² defined *via* shadow mask.

**Characterization methods**

**Current density–voltage (*J-V*) measurements**

*J–V* characteristics were measured using a class AAA solar simulator (Newport, Oriel Sol3A) at a power density of 100 mW cm⁻², calibrated using a reference silicon solar cell (Newport, calibrated 2018) equipped with a KG5 bandpass filter to simulate the AM 1.5G solar spectrum (see Figure S27). Scan rate during measurement was set to 0.6 V s⁻¹ using a source meter (Keithley, 2400 A). The stabilized PCE of PSCs was determined by tracking the MPP under continuous AM 1.5G illumination for 300 s. The solar cell temperature during measurements was actively regulated by a Peltier element connected to a microcontroller set to 25 °C.

**X-ray diffraction spectra (XRD)**

Crystal structure analysis of perovskite layers was carried out utilizing an XRD system (Bruker D2Phaser system) with Cu-$K_\alpha$ radiation (λ = 1.5405 Å) in Bragg–Brentano configuration using a LynxEye detector. XRD spectra were measured for perovskite layers deposited on substrates with unpatterned ITO and the stated HTL.

**X-ray emission spectroscopy (XES) measurements**

For XES, the samples were processed at KIT, sealed in dry nitrogen atmosphere, shipped to Beamline 8.0.1 at the Advanced Light Source (ALS), Lawrence Berkeley National Laboratory (LBNL), and inserted into the SALSA endstation[74] without any air exposure. The N K emission spectra were recorded with a high-transmission soft x-ray spectrometer,[75] and the energy axis was calibrated using BN and $CaSO_4$.[76] 2PACz reference powder (TCI AMERICA, purity >98.0%) was pressed into a pellet in a nitrogen-filled glovebox. To mitigate x-ray beam-induced changes, all samples were scanned under the x-ray beam with a speed of 600 μm s⁻¹, which corresponds to an exposure time of 50 ms for each spot. Detailed studies were performed to ascertain that beam-induced changes observed in static measurements were absent in the data with sample scanning.

**Nuclear magnetic resonance (NMR) measurements**

NMR spectra (¹H) were recorded in d6-DMSP at 400 MHz with a Bruker spectrometer at 298 K. Chemical shifts are reported in parts per million (ppm) relative to the traces of d5-DMSO ($\delta_H$= 2.50 ppm) in the corresponding fully deuterated solvent. All materials were weighed out in a nitrogen filled glovebox and mixed under ambient conditions.

**Density functional theory (DFT) calculations**

*Ab-initio computations*

Computations were performed using DFT as implemented in the quantum espresso package with PBE exchange-correlation functional.[77] The core electrons were modeled using ultrasoft pseudopotentials, and a plane wave basis set cutoff of 55 Ry was used and the dispersion correction was included at the DFT-D3 level.[78] Relaxations of the structures were performed using the conjugate gradient methods with a convergence threshold on forces set to 10$^{-3}$ (in atomic units).

*Determination of entropic contribution*

We utilized the entropy argument from Chen et al.[66] to explain the stability of α-FAPbI$_3$ above room temperature. If we assume that the formamidinium cation is geometrically restricted within the hexagonal δ-FAPbI$_3$ unit cell (in line with the X-ray analysis by Chen et al.),[66] then the Gibbs free energy of δ-FAPbI$_3$ is equal to its total internal energy, namely:

$$G^\delta = E^\delta_{DFT}$$

For the cubic α-FAPbI$_3$, the X-ray data clearly indicates a random rotation of the FA$^+$ within the cubic unit cell. Consequently, this free and random motion will result in an entropy contribution to the Gibbs free energy of the α-FAPbI$_3$:

$$G^\alpha = E^\alpha_{DFT} - T \times S$$

where *T* is the temperature and the *S* is the entropy of FA$^+$ equal to:

$$S = \frac{3}{2} k_B \times \left\{1 + \ln(0.478 \times k_B T \times (I_1 I_2 I_3)^{\frac{1}{3}})\right\}$$

In the above expression, $k_B$ and *T* are the Boltzmann constant and temperature, respectively, and the principal moments of inertia $I_i$ of FA$^+$ are 11.644, 60.161, and 71.806 *u* Å, where the unified atomic mass unit, *u*, is 1.6605 × 10$^{-27}$ kg.[66] At T = 18 °C, the entropy contribution amounts to −253 meV.

## Acknowledgements

The authors gratefully acknowledge financial support by the Initiating and Networking funding of the Helmholtz Association [HYIG of U.W.P. (VHNG-1148)], the Helmholtz Energy Materials Foundry (HEMF), the German Federal Ministry for Economics and Climate Action (BMWi) through the project 27Plus6 (03EE1056B) as well as project SHAPE (03EE1123A), funding from the European Commison, Horizon Europe research and innovation program under the NEXUS grant (101075330), and the Karlsruhe School of Optics and Photonics (KSOP). The authors acknowledge the Helmholtz Association (program-oriented funding IV, Materials and Technologies for the Energy Transition, Topic 1: Photovoltaics and Wind Energy, Code: 38.01.02). The authors also acknowledge the Karlsruhe Micro Nano Facility (KMNF), who provided access for atomic layer deposition and thank Dr. Andreas Rapp, Despina Savviduo and Tanja Ohmer-Scherrer for their support in measuring herein presented NMR data.

A.T., M.E., and G.G. gratefully acknowledge support from the Klaus Tschira Foundation, funding from the European Research Council (ERC) under the European Union's Horizon 2020 research and innovation program (Grant agreement No. 101042290 PATTERNCHEM), and the computing time provided to them on the high-performance computer Noctua 2 under the project hpc-prf-patchem at the NHR Center PC2, funded by the Federal Ministry

## Author contributions

T.F. and J.P. contributed equally to this work. T.F. conceived the initial idea for this study and developed it further with support of J.P. T.F. fabricated the presented thin-film samples, performed and analyzed JV, EQE, MPP, XRD, PLQY measurements, and analyzed TSPSC measurements. J.P. was heavily involved in sample fabrication confirming initial concepts, assisted in experimental design, and performed and analyzed all NMR measurements. A.T., M.E., and G.G. performed and analyzed all DFT calculations. D.H., C.W., C.H., and L.W. performed and analyzed all XES measurements. B.H. performed all TSPSC measurements and assisted in the fitting routine. A.D. was involved in sample fabrication confirming initial concepts, provided valuable discussions and performed all SEM measurements. P.F. and U.W.P. provided valuable discussion and supervised the project. T.F. drafted the manuscript with support of J.P., A.T., and D.H., and the manuscript was written through contributions of all authors. All authors reviewed and commented on the paper.

## Declaration of interests

The authors declare no competing interests.